\documentclass[runningheads]{llncs}
\usepackage{graphicx}
\usepackage{amssymb, amsmath, bm, latexsym,comment}
\usepackage{multirow}
\usepackage{xcolor}
\usepackage{subfigure}
\usepackage{indentfirst}
\usepackage{setspace}
\usepackage{verbatim}
\usepackage{array}

\providecommand{\Zxhreftb}[1]{Table~\ref{#1}}

\providecommand{\zxhreffig}[1]{Fig.~\ref{#1}}
\providecommand{\Zxhreffig}[1]{Fig.~\ref{#1}}

\providecommand{\citep}[1]{\cite{#1}}

\begin{document}
\title{Joint Left Atrial Segmentation and Scar Quantification Based on a DNN with Spatial Encoding and Shape Attention}
\titlerunning{Joint LA segmentation and Scar Quantification via MTL-SESA Net}

\author{Lei Li\inst{1, 2, 3} \and
Xin Weng \inst{2} \and
Julia A. Schnabel \inst{3} \and
Xiahai Zhuang\inst{2} \thanks{Corresponding author: zxh@fudan.edu.cn.}
}
\authorrunning{L. Li et al.}

\institute{School of Biomedical Engineering, Shanghai Jiao Tong University, Shanghai, China \and
School of Data Science, Fudan University, Shanghai, China \\
\email{zxh@fudan.edu.cn} \and 
School of Biomedical Engineering and Imaging Sciences, King’s College London, London, UK}

\maketitle 
\begin{abstract}
We propose an end-to-end deep neural network (DNN) which can simultaneously segment the left atrial (LA) cavity and quantify LA scars.
The framework incorporates the continuous spatial information of the target by introducing a spatially encoded (SE) loss based on the distance transform map. 
Compared to conventional binary label based loss, the proposed SE loss can reduce noisy patches in the resulting segmentation, which is commonly seen for deep learning-based methods.
To fully utilize the inherent spatial relationship between LA and LA scars, we further propose a shape attention (SA) mechanism through an explicit surface projection to build an end-to-end-trainable model.
Specifically, the SA scheme is embedded into a two-task network to perform the joint LA segmentation and scar quantification. 
Moreover, the proposed method can alleviate the severe class-imbalance problem when detecting small and discrete targets like scars. 
We evaluated the proposed framework on 60 LGE MRI data from the MICCAI2018 LA challenge.
For LA segmentation, the proposed method reduced the mean Hausdorff distance from 36.4 mm to 20.0 mm compared to the 3D basic U-Net using the binary cross-entropy loss.
For scar quantification, the method was compared with the results or algorithms reported in the literature and demonstrated better performance.

\end{abstract}

\section{Introduction}
Atrial fibrillation (AF) is the most common cardiac arrhythmia which increases the risk of stroke, heart failure and death~\cite{journal/cir/chugh2014}.
Radiofrequency ablation is a promising procedure for treating AF, where patient selection and outcome prediction of such therapy can be improved through left atrial (LA) scar localization and quantification. 
Atrial scars are located on the LA wall, thus it normally requires LA/ LA wall segmentation to exclude confounding enhanced tissues from other substructures of the heart.
Late gadolinium enhanced magnetic resonance imaging (LGE MRI) has been an important tool for scar visualization and quantification.
Manual delineations of LGE MRI can be subjective and labor-intensive.
However, automating this segmentation remains challenging, mainly due to the various LA shapes, thin LA wall, poor image quality and enhanced noise from surrounding tissues.

Limited studies have been reported in the literature to develop automatic LA segmentation and scar quantification algorithms. 
For LA segmentation, Xiong et al. proposed a dual fully convolutional neural network (CNN) \cite{journal/TMI/xiong2018}.
In an LA segmentation challenge~\cite{link/LAseg2018},
Chen et al. presented a two-task network for atrial segmentation and post/ pre classification to incorporate the prior information of the patient category \cite{conf/STACOM/chen2018}.
Nunez et al. achieved LA segmentation by combining multi-atlas segmentation and shape modeling of LA \cite{conf/STACOM/nunez2018}.
Recently, Yu et al. designed an uncertainty-aware semi-supervised framework for LA segmentation \cite{conf/MICCAI/yu2019}.
For scar quantification, most of the current works adopted threshold-based methods that relied on manual LA wall segmentation~\cite{journal/jcmr/Karim2013}.
Some other conventional algorithms, such as Gaussian mixture model (GMM)~\cite{journal/TEHM/karim2014}, also required an accurate initialization of LA or LA wall.
However, automatic LA wall segmentation is complex and challenging due to its inherent thin thickness ($1\sim2$ mm)~\cite{journal/MedAI/karim2018}.
Recent studies show that the thickness can be ignored, as clinical studies mainly focus on the location and extent of scars~\cite{journal/tmi/ravanelli2013,journal/MedAI/li2020}.
For example, Li et al. proposed a graph-cuts framework for scar quantification on the LA surface mesh, where the weights of the graph were learned via a multi-scale CNN~\cite{journal/MedAI/li2020}.
However, they did not achieve an end-to-end training, i.e., the multi-scale CNN and graph-cuts were separated into two sub-tasks.

Recently, deep learning (DL)-based methods have achieved promising performance for cardiac image segmentation.
However, most DL-based segmentation methods are trained with a loss only considering a label mask in a discrete space.
Due to the lack of spatial information, predictions commonly tend to be blurry in boundary, and it leads to noisy segmentation with large outliers.
To solve this problem, several strategies have been employed, such as graph-cuts/ CRF regularization~\cite{journal/MedAI/li2020,journal/MedAI/kamnitsas2017}, and deformation combining shape priors~\cite{conf/MICCAI/zeng2019}.


\begin{figure*}[t]\center
 \includegraphics[width=0.98\textwidth]{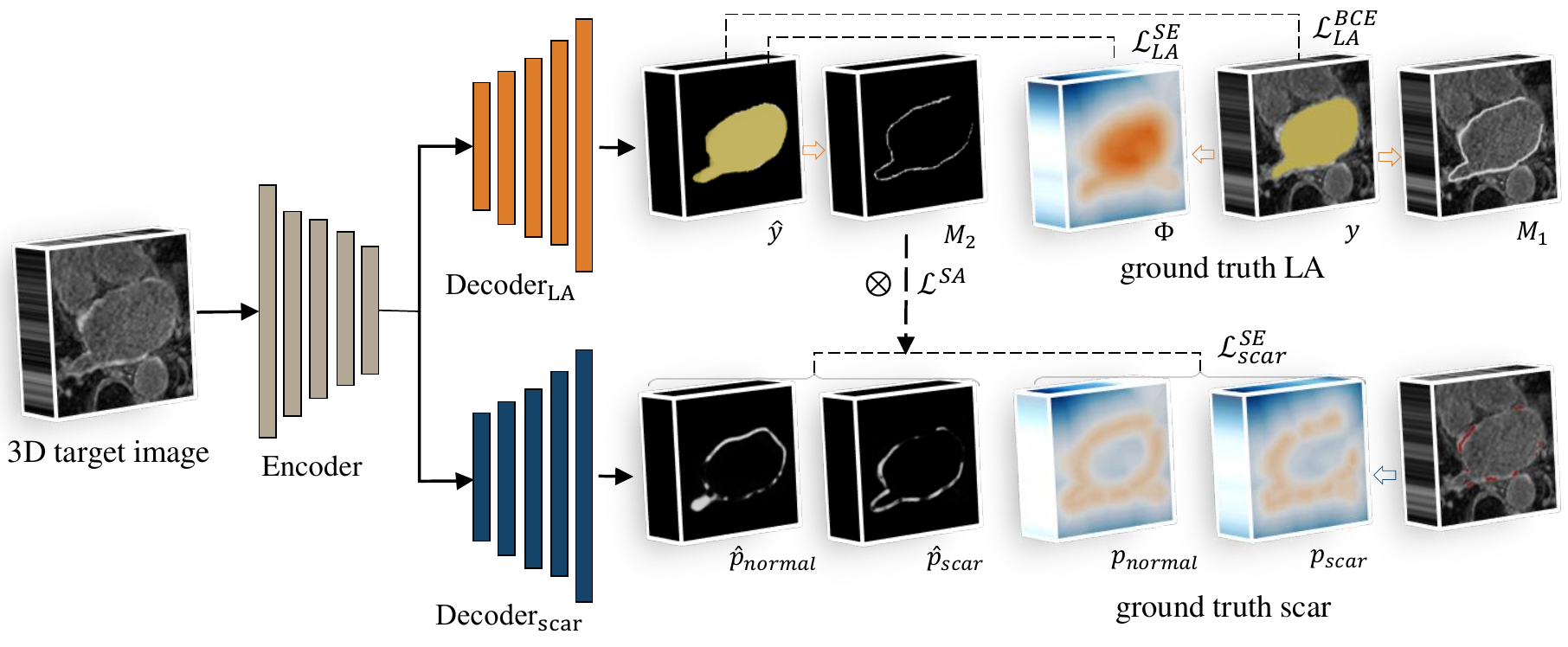}\\[-2ex]
   \caption{The proposed MTL-SESA network for joint LA segmentation and scar quantification. Note that the skip connections between the encoder and two decoders are omitted here.}
\label{fig:method:network}\end{figure*}

In this work, we present an end-to-end multi-task learning network for joint LA segmentation and scar quantification.
The proposed method incorporates spatial information in the pipeline to eliminate outliers for LA segmentation, with additional benefits for scar quantification. 
This is achieved by introducing a spatially encoded loss based on the distance transform map, without any modifications of the network.
To utilize the spatial relationship between LA and scars, we adopt the LA boundary as an attention mask on the scar map, namely surface projection, to achieve shape attention. 
Therefore, an end-to-end learning framework is created for simultaneous LA segmentation, scar projection and quantification via the multi-task learning (MTL) network embedding the spatial encoding (SE) and boundary shape attention (SA), namely MTL-SESA network.

\section{Method} \label{method}
\zxhreffig{fig:method:network} provides an overview of the proposed framework.
The proposed network is a modified U-Net consisting of two decoders for LA segmentation and scar quantification, respectively.
In Section \ref{method:LA}, a SE loss based on the distance transform map is introduced as a regularization term for LA segmentation.
For scar segmentation, a SE loss based on the distance probability map is employed, followed by a spatial projection (see Section \ref{method:scar}).
Section \ref{method:multi_task} presents the specific SA scheme embedded in the MTL network for the predictions of LA and LA scars in an end-to-end style.

\subsection{Spatially Encoded Constraint for LA Segmentation}\label{method:LA}
A SE loss based on the signed distance transform map (DTM) is employed as a regularization term to represent a spatial vicinity to the target label. 
Given a target label, the signed DTM for each pixel $x_i$ can be defined as:
\begin{equation}
\phi(x_i)=\begin{cases} -d^\beta & x_i \in \Omega_{in} \\0 & x_i \in S\\ d^\beta & x_i \in \Omega_{out} \end{cases} 
\end{equation}
where $\Omega_{in}$ and $\Omega_{out}$ respectively indicate the region inside and outside the target label, $S$ denotes the surface boundary, $d$ represents the distance from pixel $x_i$ to the nearest point on $S$, and $\beta$ is a hyperparameter.
The binary cross-entropy (BCE) loss and the additional SE loss for LA segmentation can be defined as:
\begin{equation}
  \mathcal L_{LA}^{BCE} = \sum_{i=1}^N y_i \cdot log(\hat{y}(x_i; \theta)) + (1-y_i) \cdot log(1-\hat{y}(x_i; \theta))
\end{equation}
\begin{equation}
  \mathcal L_{LA}^{SE} = \sum_{i=1}^N (\hat{y}(x_i; \theta)-0.5) \cdot \phi(x_i)
\end{equation}
where $\hat{y}$ and $y$ ($y\in\{0,1\}$) are the prediction of LA and its ground truth, respectively, and $\cdot$ denotes element-wise product.

\subsection{Spatially Encoded Constraint with an Explicit Projection for Scar Quantification}  \label{method:scar}
For scar quantification, we encode the spatial information by adopting the distance probability map of normal wall and scar region as the ground truth instead of binary scar label.
This is because the scar region can be very small and discrete, thus its detection presents significant challenges to current DL-based methods due to the class-imbalance problem.
In contrast to traditional DL-based algorithms optimizing in a discrete space, the distance probability map considers the continuous spatial information of scars.
Specifically, we separately obtain the DTM of the scar and normal wall from a manual scar label, and convert both into probability maps $p(x_i)= [p_{normal}, p_{scar}]$.
Here $p=e^{-d'}$ and $d'$ is the nearest distance to the boundary of normal wall or scar for pixel $x_i$.
Then, the SE loss for scar quantification can be defined as:
\begin{equation}
  \mathcal L_{scar}^{SE} = \sum_{i=1}^N (\hat{p}(x_i; \theta) - p(x_i))^2
\end{equation}
where $\hat{p}$ ($\hat{p} = [\hat{p}_{normal}, \hat{p}_{scar}]$) is the predicted distance probability map of both normal wall and scar region.
Note that the situation of $\hat{p}_{normal} + \hat{p}_{scar} > 1$ sometimes exists.
One can compare these two probabilities to extract scars instead of employing a fixed threshold.

To ignore the wall thickness which varies from different positions and patients~\cite{journal/MedAI/karim2018}, the extracted scars are explicitly projected onto the LA surface.
Therefore, the volume-based scar segmentation is converted into a surface-based scar quantification through the spatially explicit projection.
However, the pixel-based classification in the surface-based quantification task only includes very limited information, i.e., the intensity value of one pixel. 
In contrast to extracting multi-scale patches along the LA surface~\cite{journal/MedAI/li2020}, we employ the SE loss to learn the spatial features near the LA surface.
Similar to~\cite{journal/MedAI/li2020}, the SE loss can also be beneficial to improving the robustness of the framework against the LA segmentation errors.

\subsection{Multi-task Learning with an End-to-end Trainable Shape Attention} \label{method:multi_task}
To employ the spatial relationship between LA and atrial scars, we design an MTL network including two decoders, i.e., one for LA and the other for scar segmentation.
As \zxhreffig{fig:method:network} shows, the Decoder$_\mathrm{LA}$ is supervised by $\mathcal L_{LA}^{BCE}$ and $\mathcal L_{LA}^{SE}$, and the Decoder$_\mathrm{scar}$ is supervised by $\mathcal L_{scar}^{SE}$.
To explicitly learn the relationship between the two tasks, we extract the LA boundary from the predicted LA as an attention mask for the training of Decoder$_\mathrm{scar}$, namely explicit projection mentioned in Section~\ref{method:scar}.
An SA loss is introduced to enforce the attention of Decoder$_\mathrm{scar}$ on the LA boundary:
\begin{equation}
  \mathcal L^{SA} = \sum_{i=1}^N (M \cdot (\Delta \hat{p}(x_i; \theta) - \Delta p(x_i)))^2
\end{equation}
where $\Delta \hat{p} = \hat{p}_{normal} - \hat{p}_{scar}$, 
$\Delta p = p_{normal}-p_{scar}$, and $M$ is the boundary attention mask, which can be generated from the gold standard segmentation of LA ($M_{1}$) as well as the predicted LA ($M_{2}$).
Hence, the total loss of the framework is defined by combining all the losses mentioned above:
\begin{equation}
   \mathcal L = \mathcal L_{LA}^{BCE} + \lambda_{LA}\mathcal L_{LA}^{SE} + \lambda_{scar}\mathcal L_{scar}^{SE}
        + \lambda_{M_{1}}\mathcal L^{SA}_{scarM_1} + \lambda_{M_{2}}\mathcal L^{SA}_{scarM_2}
\end{equation}
where $\lambda_{LA}$, $\lambda_{scar}$, $\lambda_{M_{1}}$ and $\lambda_{M_{2}}$ are balancing parameters.

\section{Experiments}

\subsection{Materials}
\subsubsection{Data Acquisition and Pre-processing.}
The data is from the MICCAI2018 LA challenge~\cite{link/LAseg2018}. 
The 100 LGE MRI training data, with manual segmentation of LA, consists of 60 post-ablation and 40 pre-ablation data.
In this work, we chose the 60 post-ablation data for manual segmentation of the LA scars and employed them for experiments.
The LGE MRIs were acquired with a resolution of $0.625\!\times\!0.625\!\times\!0.625$ mm and reconstructed to $1\!\times\!1\!\times\!1$ mm.
All images were cropped into a unified size of $208\!\times\!208\!\times\!80$ centering at the heart region and were normalized using Z-score.
We split the images into two sets, i.e., one with 40 images for training and the other with 20 for the test. 

\subsubsection{Gold Standard and Evaluation.}
The challenge provides LA manual segmentation for the training data, and scars of the 60 post-ablation data were manually delineated by a well-trained expert.
These manual segmentations were considered as the gold standard.
For LA segmentation evaluation, Dice volume overlap, average surface distance (ASD) and Hausdorff distance (HD) were applied.
For scar quantification evaluation, the manual and (semi-) automatic segmentation results were first projected onto the manually segmented LA surface.
Then, the \emph{Accuracy} measurement of the two areas in the projected surface, Dice of scars (Dice$_\mathrm{scar}$) and generalized Dice score ($G$Dice) were used as indicators of the accuracy of scar quantification. 

\subsubsection{Implementation.}
The framework was implemented in PyTorch, running on a computer with 1.90 GHz Intel(R) Xeon(R) E5-2620 CPU and an NVIDIA TITAN X GPU. 
We used the SGD optimizer to update the network parameters (weight decay=0.0001, momentum=0.9). 
The initial learning rate was set to 0.001 and divided by 10 every 4000 iterations. 
The balancing parameters in Section \ref{method:multi_task}, were set as follows, $\lambda_{LA}=0.01$, $\lambda_{scar}=10$, $\lambda_{M_{1}}=0.01$ and $\lambda_{M_{2}}=0.001$, where $\lambda_{LA}$ and $\lambda_{M_{2}}$ was multiplied by 1.1 every 200 iterations.
The inference of the networks required about 8 seconds to process one test image.

\subsection{Result}

\begin{figure*}[!t]\center
	\subfigure[] {\includegraphics[width=0.47\textwidth]{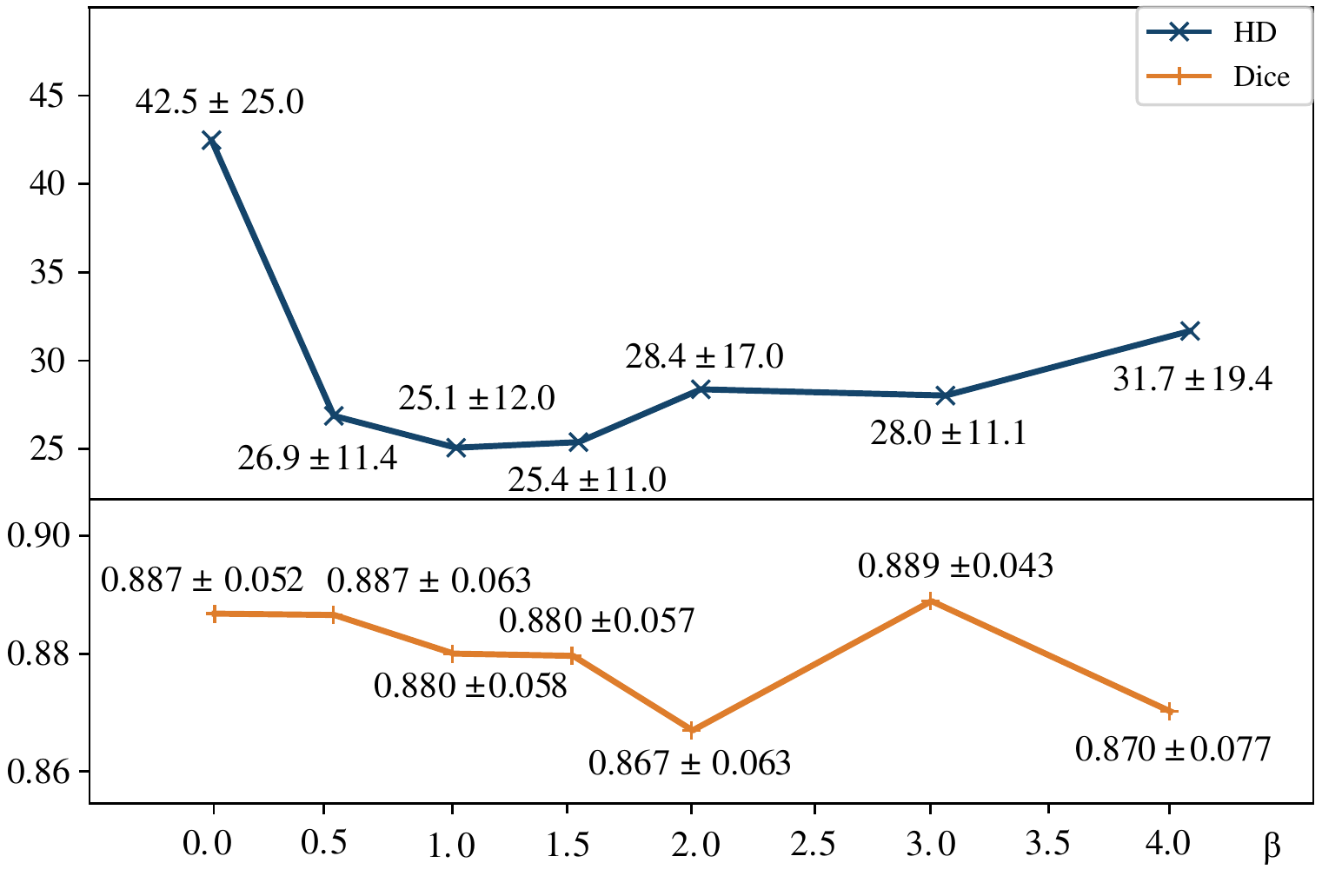}}
	\subfigure[] {\includegraphics[width=0.51\textwidth]{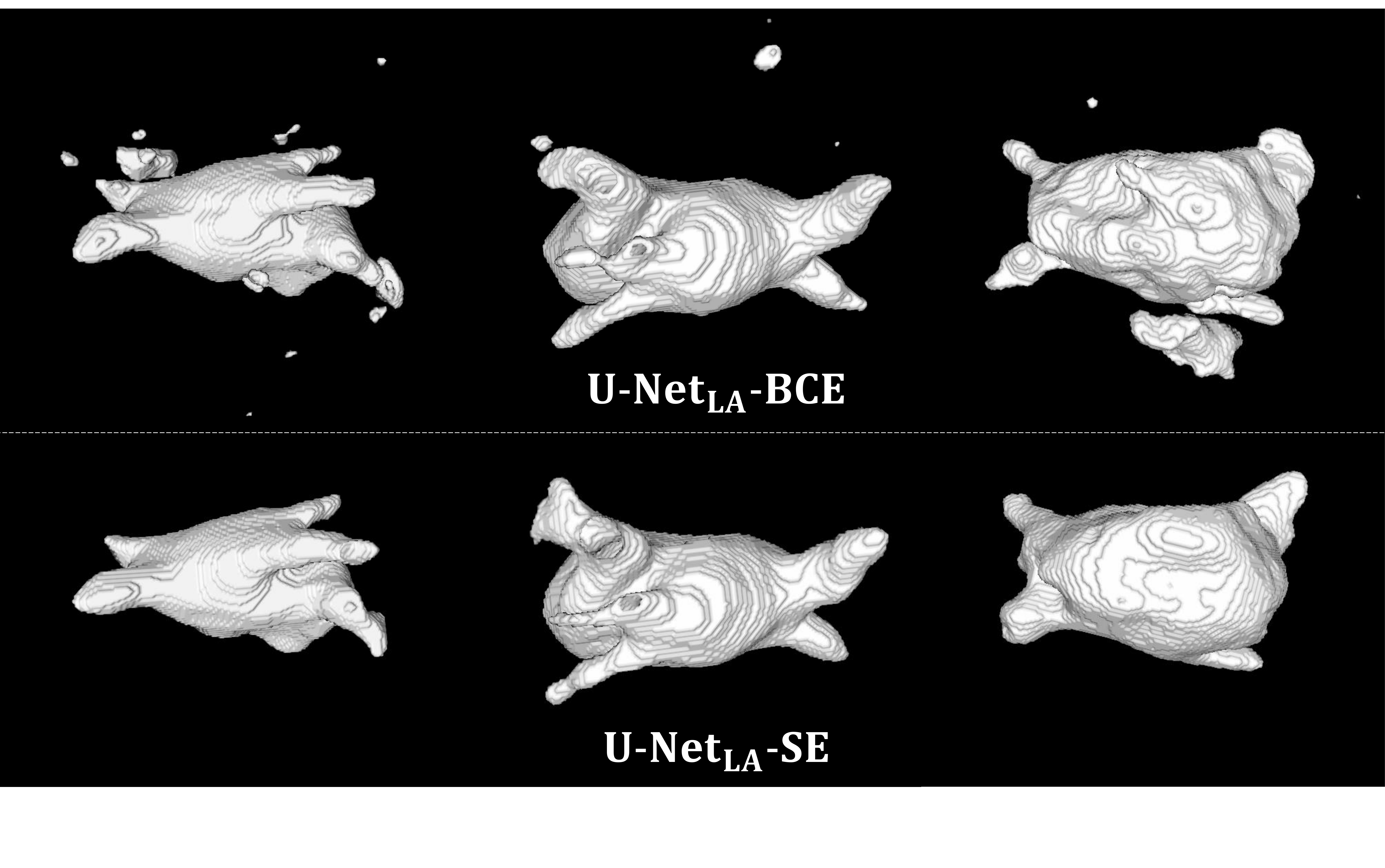}}
	\caption{
		Quantitative and qualitative evaluation results of the proposed SE loss for LA segmentation: (a) Dice and HD of the LA segmentation results after combining the SE loss, i.e., U-Net$_\mathrm{LA}$-SE with different $\beta$ for DTM; 
		(b) 3D visualization of the LA segmentation results of three typical cases by U-Net$_\mathrm{LA}$-BCE and U-Net$_\mathrm{LA}$-SE.}
	\label{fig:result:LA_SE}\end{figure*}

\begin{table*} [t] \center
    \caption{
    Summary of the quantitative evaluation results of LA segmentation. Here, U-Net$_\mathrm{LA}$ uses the original U-Net architecture for LA segmentation; 
    MTL means that the methods are based on the architecture in \zxhreffig{fig:method:network} with two decoders; BCE, SE, SA and SESA refer to the different loss functions.
    The proposed method is denoted as MTL-SESA.
     }
\label{tb:result:LA}
{\small
\begin{tabular}{ l| l *{5}{@{\ \,} l }}\hline
Method    & \quad Dice  & \quad ASD (mm) & \quad HD (mm) \\
\hline
U-Net$_\mathrm{LA}$-BCE           &$ 0.889 \pm 0.035 $&  \quad $ 2.12 \pm 0.80 $&  \quad $ 36.4 \pm 23.6 $\\
U-Net$_\mathrm{LA}$-SE            &$ 0.880 \pm 0.058 $&  \quad $ 2.36 \pm 1.49 $&  \quad $ 25.1 \pm 11.9 $\\
\hline\hline
MTL-BCE                           &$ 0.890 \pm 0.042 $&  \quad $ 2.11 \pm 1.01 $&  \quad $ 28.5 \pm 14.0 $\\
MTL-SE                            &$ 0.909 \pm 0.033 $&  \quad $ 1.69 \pm 0.69 $&  \quad $ 22.4 \pm 9.80 $\\
MTL-SESA                          &\bm{$ 0.913 \pm 0.032 $}&  \quad \bm{$ 1.60 \pm 0.72 $}& \quad \bm{$ 20.0 \pm 9.59 $}\\
\hline
\end{tabular} }\\
\end{table*}

\begin{table*} [t] \center
    \caption{                                                                                                         
    Summary of the quantitative evaluation results of scar quantification. 
    Here, LA$_\mathrm{M}$ denotes that scar quantification is based on the manually segmented LA, while LA$_{\mathrm{U\mbox{-}Net}}$ indicates that it is based on the U-Net$_\mathrm{LA}$-BCE segmentation;
    U-Net$_\mathrm{scar}$ is the scar segmentation directly based on the U-Net architecture with different loss functions; 
    The inter-observer variation (Inter-Ob) is calculated from randomly selected twelve subjects.
     }
\label{tb:result:scar}
{\small
\begin{tabular}{ l| l *{4}{@{\ \,} l }}\hline
Method       & \quad Accuracy & \qquad Dice$_\mathrm{scar}$ & \qquad $G$Dice\\
\hline
LA$_\mathrm{M}$+Otsu~\cite{journal/tmi/ravanelli2013} &$ 0.750 \pm 0.219 $& \quad $ 0.420 \pm 0.106 $ & \quad $ 0.750 \pm 0.188 $\\
LA$_\mathrm{M}$+MGMM~\cite{journal/TBME/liu2017}      &$ 0.717 \pm 0.250 $& \quad $ 0.499 \pm 0.148 $ & \quad $ 0.725 \pm 0.239 $\\
LA$_\mathrm{M}$+LearnGC~\cite{journal/MedAI/li2020}   &$ 0.868 \pm 0.024 $& \quad $ 0.481 \pm 0.151 $ & \quad $ 0.856 \pm 0.029 $\\ 
\hline
LA$_{\mathrm{U\mbox{-}Net}}$+Otsu  &$ 0.604 \pm 0.339 $& \quad $ 0.359 \pm 0.106 $ & \quad $ 0.567 \pm 0.359 $ \\
LA$_{\mathrm{U\mbox{-}Net}}$+MGMM  &$ 0.579 \pm 0.334 $& \quad $ 0.430 \pm 0.174 $ & \quad $ 0.556 \pm 0.370 $ \\
\hline
U-Net$_\mathrm{scar}$-BCE          &$ 0.866 \pm 0.032 $& \quad $ 0.357 \pm 0.199 $ & \quad $ 0.843 \pm 0.043 $\\
U-Net$_\mathrm{scar}$-Dice         &$ 0.881 \pm 0.030 $& \quad $ 0.374 \pm 0.156 $ & \quad $ 0.854 \pm 0.041 $\\
U-Net$_\mathrm{scar}$-SE           &$ 0.868 \pm 0.026 $& \quad $ 0.485 \pm 0.129 $ & \quad $ 0.863 \pm 0.026 $\\
\hline\hline
MTL-BCE                            &\bm{$ 0.887 \pm 0.023 $}& \quad $ 0.484 \pm 0.099 $ & \quad \bm{$ 0.872 \pm 0.024 $}\\
MTL-SE                             &$ 0.882 \pm 0.026 $& \quad $ 0.518 \pm 0.110 $ & \quad $ 0.871 \pm 0.024 $\\
MTL-SESA                           &$ 0.867 \pm 0.032 $& \quad \bm{$ 0.543 \pm 0.097 $} & \quad $ 0.868 \pm 0.028 $\\
\hline\hline
Inter-Ob                           &$ 0.891 \pm 0.017 $& \quad $ 0.580 \pm 0.110 $ & \quad $ 0.888 \pm 0.022 $\\
\hline
\end{tabular} }\\
\end{table*}
	
\begin{figure*}[t]\center
	\includegraphics[width=1.0\textwidth]{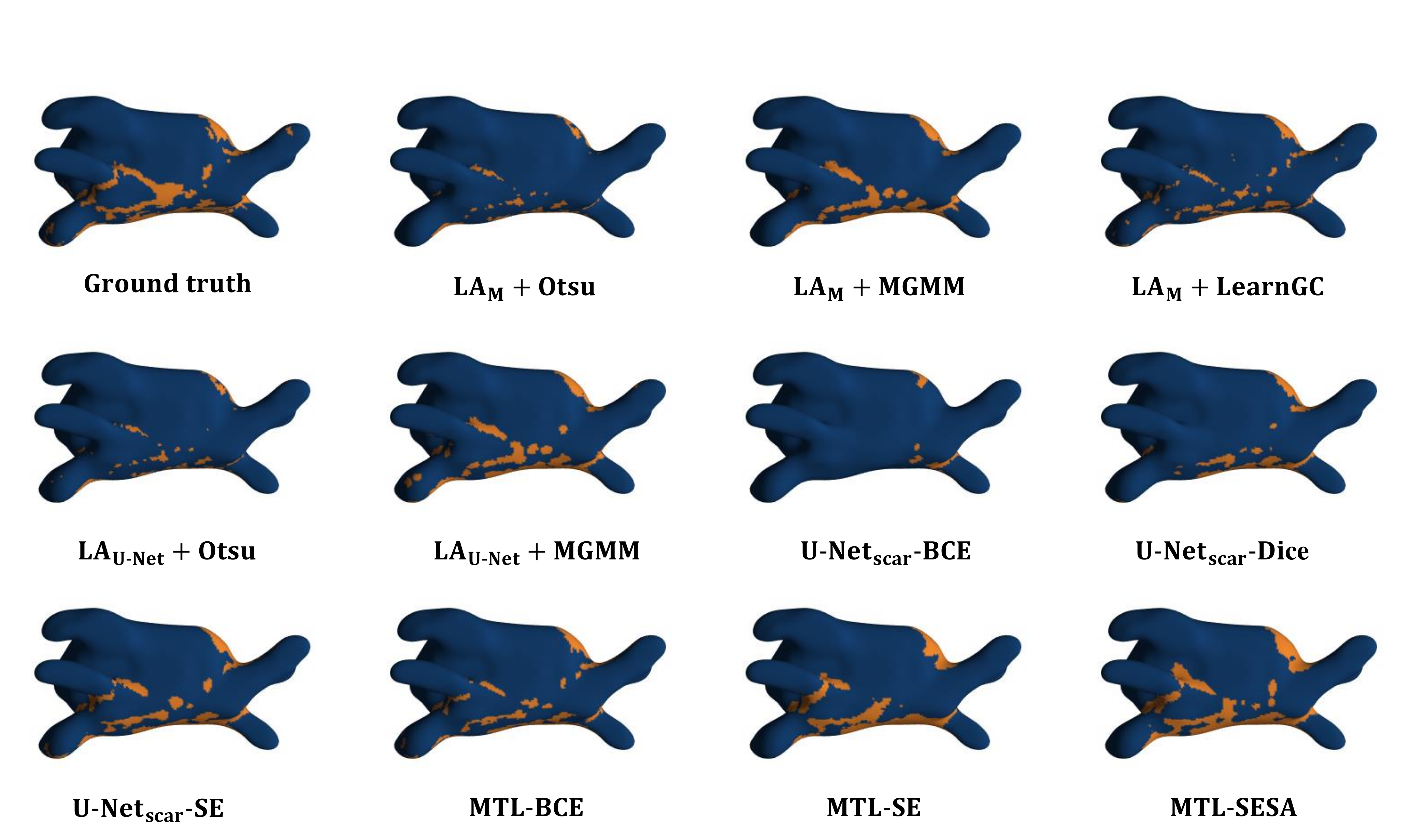}\\[-2ex]
	\caption{3D visualization of the LA scar localization by the eleven methods. The scarring areas are labeled in orange on the LA surface, which is constructed from LA$_\mathrm{M}$ labeled in blue. 
	}
	\label{fig:result:3d_results}\end{figure*}

\subsubsection{Parameter Study.}
To explore the effectiveness of the SE loss, we compared the results of the proposed scheme for LA segmentation using different values of $\beta$ for DTM in Eq. (1).
\Zxhreffig{fig:result:LA_SE} (a) provides the results in terms of Dice and HD,
and \Zxhreffig{fig:result:LA_SE} (b) visualizes three examples for illustrating the difference of the results using or without using the SE loss.
One can see that with the SE loss, U-Net$_\mathrm{LA}$-SE evidently reduced clutter and disconnected parts in the segmentation compared to U-Net$_\mathrm{LA}$-BCE, and significantly improved the HD of the resulting segmentation ($p<0.001$), though the Dice score may not be very different.
Also, U-Net$_\mathrm{LA}$-SE showed stable performance with different values of $\beta$ except for too extreme values.
In the following experiments, $\beta$ was set to 1.

\subsubsection{Ablation Study.}
\Zxhreftb{tb:result:LA} and \Zxhreftb{tb:result:scar} present the quantitative results of different methods for LA segmentation and scar quantification, respectively.
For LA segmentation, combining the proposed SE loss performed better than only using the BCE loss.
For scar quantification, the SE loss also showed promising performance compared to the conventional losses in terms of Dice$_\mathrm{scar}$.
LA segmentation and scar quantification both benefited from the proposed MTL scheme comparing to achieving the two tasks separately.
The results were further improved after introducing the newly-designed SE and SA loss in terms of Dice$_\mathrm{scar}$ ($p\leq0.001$), but with a slightly worse \emph{Accuracy} ($p\leq0.001$) and $G$Dice ($p>0.1$).
\Zxhreffig{fig:result:3d_results} visualizes an example for illustrating the segmentation and quantification results of scars from the mentioned methods in \Zxhreftb{tb:result:scar}.
Compared to U-Net$_\mathrm{scar}$-BCE and U-Net$_\mathrm{scar}$-Dice, MTL-BCE improved the performance, thanks to the MTL network architecture.
When the proposed SE and SA loss were included, some small and discrete scars were also detected, and an end-to-end scar quantification and projection was achieved.

\subsubsection{Comparisons with Literature.}
\Zxhreftb{tb:result:scar} and \Zxhreffig{fig:result:3d_results} also present the scar quantification results from some state-of-the-art algorithms, i.e., Otsu~\cite{journal/tmi/ravanelli2013}, multi-component GMM (MGMM)~\cite{journal/TBME/liu2017}, LearnGC~\cite{journal/MedAI/li2020} and U-Net$_\mathrm{scar}$ with different loss functions.
The three (semi-) automatic methods generally obtained acceptable results, but relied on an accurate initialization of LA.
LearnGC had a similar result compared to MGMM in Dice$_\mathrm{scar}$ based on LA$_\mathrm{M}$, but its \emph{Accuracy} and $G$Dice were higher.
The proposed method performed much better than all the automatic methods in terms of Dice$_\mathrm{scar}$ with statistical significance ($p\leq0.001$).
In \Zxhreffig{fig:result:3d_results}, one can see that Otsu and U-Net$_\mathrm{scar}$ tended to under-segment the scars.
Though including Dice loss could alleviate the class-imbalance problem, it is evident that the SE loss could be more effective, which is consistent with the quantitative results in \Zxhreftb{tb:result:scar}.
MGMM and LearnGC both detected most of the scars, but LearnGC has the potential advantage of small scar detection.
The proposed method could also detect small scars and obtained a smoother segmentation result. 

\section{Conclusion}
In this work, we have proposed an end-to-end learning framework for simultaneous LA segmentation and scar quantification by combining the SE and SA loss.
The proposed algorithm has been applied to 60 image volumes acquired from AF patients and obtained comparable results to inter-observer variations.
The results have demonstrated the effectiveness of the proposed SE and SA loss, and showed the superiority of segmentation performance over the conventional schemes.
Particularly, the proposed SE loss substantially reduced the outliers, which frequently occurs in the prediction of DL-based methods.
Our technique can be easily extended to other segmentation tasks, especially for discrete and small targets such as lesions.
A limitation of this work is that the gold standard was constructed from the manual delineation of only one expert.
Besides, the target included in this study is only post-ablation AF patients. 
In future work, we will combine multiple experts to construct the gold standard, and consider both pre- and post-ablation data.

\subsubsection{Acknowledgement.}
This work was supported by the National Natural Science Foundation of China (61971142), and L. Li was partially supported by the CSC Scholarship.

\bibliographystyle{splncs04}
\bibliography{AllBibliography_MICCAI2020}

\begin{thebibliography}{10}
\providecommand{\url}[1]{\texttt{#1}}
\providecommand{\urlprefix}{URL }
\providecommand{\doi}[1]{https://doi.org/#1}

\bibitem{conf/STACOM/chen2018}
Chen, C., Bai, W., Rueckert, D.: Multi-task learning for left atrial
  segmentation on {GE-MRI}. In: International Workshop on Statistical Atlases
  and Computational Models of the Heart. pp. 292--301. Springer (2018)

\bibitem{journal/cir/chugh2014}
Chugh, S.S., Havmoeller, R., Narayanan, K., Singh, D., Rienstra, M., Benjamin,
  E.J., Gillum, R.F., Kim, Y.H., McAnulty~Jr, J.H., Zheng, Z.J., et~al.:
  Worldwide epidemiology of atrial fibrillation: a global burden of disease
  2010 study. Circulation  \textbf{129}(8),  837--847 (2014)

\bibitem{journal/MedAI/kamnitsas2017}
Kamnitsas, K., Ledig, C., Newcombe, V.F., Simpson, J.P., Kane, A.D., Menon,
  D.K., Rueckert, D., Glocker, B.: Efficient multi-scale 3{D} {CNN} with fully
  connected {CRF} for accurate brain lesion segmentation. Medical image
  analysis  \textbf{36},  61--78 (2017)

\bibitem{journal/TEHM/karim2014}
Karim, R., Arujuna, A., Housden, R.J., Gill, J., Cliffe, H., Matharu, K., Gill,
  J., Rindaldi, C.A., O'Neill, M., Rueckert, D., et~al.: A method to
  standardize quantification of left atrial scar from delayed-enhancement {MR}
  images. IEEE journal of translational engineering in health and medicine
  \textbf{2},  1--15 (2014)

\bibitem{journal/MedAI/karim2018}
Karim, R., Blake, L.E., Inoue, J., Tao, Q., Jia, S., Housden, R.J., Bhagirath,
  P., Duval, J.L., Varela, M., Behar, J.M., et~al.: Algorithms for left atrial
  wall segmentation and thickness--evaluation on an open-source {CT} and {MRI}
  image database. Medical image analysis  \textbf{50},  36--53 (2018)

\bibitem{journal/jcmr/Karim2013}
Karim, R., Housden, R.J., Balasubramaniam, M., Chen, Z., Perry, D., Uddin, A.,
  Al-Beyatti, Y., Palkhi, E., Acheampong, P., Obom, S., et~al.: Evaluation of
  current algorithms for segmentation of scar tissue from late gadolinium
  enhancement cardiovascular magnetic resonance of the left atrium: an
  open-access grand challenge. Journal of Cardiovascular Magnetic Resonance
  \textbf{15}(1), ~105 (2013)

\bibitem{journal/MedAI/li2020}
Li, L., Wu, F., Yang, G., Xu, L., Wong, T., Mohiaddin, R., Firmin, D., Keegan,
  J., Zhuang, X.: Atrial scar quantification via multi-scale {CNN} in the
  graph-cuts framework. Medical Image Analysis  \textbf{60},  101595 (2020)

\bibitem{journal/TBME/liu2017}
Liu, J., Zhuang, X., Wu, L., An, D., Xu, J., Peters, T., Gu, L.: Myocardium
  segmentation from {DE MRI} using multicomponent {G}aussian mixture model and
  coupled level set. IEEE Transactions on Biomedical Engineering
  \textbf{64}(11),  2650--2661 (2017)

\bibitem{conf/STACOM/nunez2018}
Nu{\~n}ez-Garcia, M., Zhuang, X., Sanroma, G., Li, L., Xu, L., Butakoff, C.,
  Camara, O.: Left atrial segmentation combining multi-atlas whole heart
  labeling and shape-based atlas selection. In: International Workshop on
  Statistical Atlases and Computational Models of the Heart. pp. 302--310.
  Springer (2018)

\bibitem{journal/tmi/ravanelli2013}
Ravanelli, D., dal Piaz, E.C., Centonze, M., Casagranda, G., Marini, M.,
  Del~Greco, M., Karim, R., Rhode, K., Valentini, A.: A novel skeleton based
  quantification and 3-{D} volumetric visualization of left atrium fibrosis
  using late gadolinium enhancement magnetic resonance imaging. IEEE
  transactions on medical imaging  \textbf{33}(2),  566--576 (2013)

\bibitem{journal/TMI/xiong2018}
Xiong, Z., Fedorov, V.V., Fu, X., Cheng, E., Macleod, R., Zhao, J.: Fully
  automatic left atrium segmentation from late gadolinium enhanced magnetic
  resonance imaging using a dual fully convolutional neural network. IEEE
  transactions on medical imaging  \textbf{38}(2),  515--524 (2018)

\bibitem{conf/MICCAI/yu2019}
Yu, L., Wang, S., Li, X., Fu, C.W., Heng, P.A.: Uncertainty-aware
  self-ensembling model for semi-supervised 3{D} left atrium segmentation. In:
  International Conference on Medical Image Computing and Computer-Assisted
  Intervention. pp. 605--613. Springer (2019)

\bibitem{conf/MICCAI/zeng2019}
Zeng, Q., Karimi, D., Pang, E.H., Mohammed, S., Schneider, C., Honarvar, M.,
  Salcudean, S.E.: Liver segmentation in magnetic resonance imaging via mean
  shape fitting with fully convolutional neural networks. In: International
  Conference on Medical Image Computing and Computer-Assisted Intervention. pp.
  246--254. Springer (2019)

\bibitem{link/LAseg2018}
Zhao, J., Xiong, Z.: 2018 atrial segmentation challenge.
  \url{http://atriaseg2018.cardiacatlas.org/} (2018)

\end{thebibliography}

\end{document}